\pdfoutput=1
\documentclass[aps,10pt,nolongbibliography,prl,tightenlines,twocolumn,twoside,showpacs,superscriptaddress
]{revtex4-1}
\usepackage[caption=false]{subfig}

\usepackage[hidelinks]{hyperref}
\usepackage{color}
\usepackage{graphicx}
\usepackage{siunitx}
\usepackage{amsmath,amssymb,amsfonts}
\usepackage{bm}
\usepackage{hyperref}
\usepackage{lineno}
\usepackage{changepage}   
\usepackage{slashed}
\usepackage{float}

\usepackage[normalem]{ulem}  

\newcommand\sect[1]{\textit{#1.}---}

\def \be {\begin{align} }
	\def \ee {\end{align}}
\def \bes {\begin{subaligns} }
	\def \ees {\end{subaligns}}

\def \no {\nonumber}
\def \a {\alpha}
\def \b {\beta}

\def \d {\delta}
\def \e {\epsilon}
\def \g {\gamma}

\def \o {\omega}

\def \r {\rho}
\def \l {\lambda}
\def \m {\mu}
\def \n {\nu}
\def \s {\sigma}

\def \pd {\partial}

\def \<{\langle}
\def \>{\rangle}
\def \+{\dagger}

\def \[{\left[}
\def \]{\right]}

\newcommand{\td}[1]{\tilde{#1}}
\newcommand{\mf}[1]{\mathfrak{#1}}

\def \vp {\bm{p}}
\def \vq {\bm{q}}
\def \vx {\bm{x}}

\def \vk {\bm{k}}

\def \ve {\varepsilon}
\def \D {\Delta}
\def \G {\Gamma}

\def \<{\langle}
\def \>{\rangle}
\def \+{\dagger}

\def \[{\left[}
\def \]{\right]}

\def \srho 
{{\tilde \rho}}

\renewcommand{\sout}{\bgroup \color{red} \ULdepth=-.5ex \ULset}

\begin{document}
	\title{Hypersurface covariance in relativistic local equilibrium}
	
	\author{You-Yu Li
	}
	\email{lyouyu@hnu.edu.cn}
	\affiliation{School of Physics and Electronics, Hunan University, Changsha 410082, China} 
	\affiliation{Hunan Provincial Key Laboratory of High-Energy Scale Physics
		and Applications, Hunan University, Changsha 410082, China}
	\author{Shuai Y.\,F.~Liu
	}
	\email[Corresponding author.\\]{lshphy@hnu.edu.cn}
	\affiliation{School of Physics and Electronics, Hunan University, Changsha 410082, China} 
	\affiliation{Hunan Provincial Key Laboratory of High-Energy Scale Physics
		and Applications, Hunan University, Changsha 410082, China}

	\date{\today}
	
\begin{abstract}
	While global equilibrium is hypersurface covariant, calculations involving local equilibrium can exhibit hypersurface dependence. 
	We use the Zubarev response approach to local polarization in heavy-ion collisions as an example to show that the hypersurface dependence of the local-equilibrium part of the density operator is canceled by an overlooked nondissipative contribution from its so-called ``dissipative part''. This establishes a hypersurface-covariant formulation for general geometries.
	The remaining frame dependence arises from the medium rest frame entering the particle distribution and the time direction of the dissipative process, which are physical and usually coincide. Using a damped density-density correlator, we derive an explicit formula describing how polarization carries initial-hypersurface memory at short times and relaxes exponentially toward the medium-rest-frame result at late times. These results resolve, in principle, the longstanding conceptual disagreement over hypersurface dependence among approaches to the widely discussed sign puzzles of $\Lambda$-hyperon polarization in heavy-ion collisions.
\end{abstract}
	
\maketitle

\sect{Introduction}
The hot and dense matter created in relativistic heavy-ion collisions can mimic extreme conditions in the early Universe and provides an important platform for studying relativistic statistical physics. With particle spectra calculated using the covariant ``Cooper--Frye" formulation~\cite{PhysRevD.10.186}, statistical descriptions have successfully reproduced a range of observables~\cite{Andronic:2005yp,Andronic:2017pug}, providing a baseline for studying nonequilibrium effects~\cite{Greco:2003xt,Ravagli:2007xx,Fries:2025jfi,He:2019vgs}.

In recent years, theoretically predicted spin observables, such as $\Lambda$-hyperon polarization and vector-meson spin alignment, have been observed experimentally~\cite{Liang:2004ph,Liang:2004xn,STAR:2017ckg,STAR:2018gyt,Niida:2018hfw,Adam:2019srw,ALICE:2021pzu,STAR:2022fan,STAR:2008lcm,ALICE:2019aid,Singha:2020qns,ALICE:2022dyy}, adding a new dimension to the study of relativistic statistical physics. Unlike observables such as hadron yields and momentum spectra, which are largely determined by flow, temperature, and chemical potential, spin observables are particularly sensitive to gradients of these hydrodynamic variables. Theoretical studies of spin polarization have revealed several conceptual ambiguities in the description of local equilibrium. One concerns the pseudo-gauge dependence of the energy-momentum tensor~\cite{Becattini:2018duy}; another concerns the hypersurface dependence of the definition of local equilibrium~\cite{Liu:2021uhn,Fu:2021pok,Becattini:2021suc,Becattini:2021iol}.

The pseudo-gauge dependence of the spin tensor has received considerable attention in studies of spin hydrodynamics~\cite{Florkowski:2017ruc,Hattori:2019lfp,Bhadury:2020puc,Fukushima:2020ucl,Gallegos:2021bzp,Hongo:2021ona,Weickgenannt:2022zxs,Weickgenannt:2022jes,Montenegro:2026phf}, with several notable conceptual advances~\cite{Li:2020eon,Becattini:2025twu}. 
For the hypersurface problem, the medium rest frame provides a natural  choice~\cite{Liu:2021uhn,Li:2025pef,Hu:2021lnx,Harutyunyan:2021rmb,Dey:2024hhc,She:2024rnx,Tiwari:2024trl,Harutyunyan:2025fgs,Fang:2024vds}. Other studies preserve explicit dependence on a reference direction, represented by a vector such as $t^\mu$ or $n^\mu$ normal to a given hypersurface~\cite{Becattini:2021suc,Becattini:2021iol,Liu:2021nyg,Sheng:2024pbw,Zhang:2024mhs,Yang:2024fkn,Sheng:2025cjk,Palermo:2025imv}, such as along the time axis of the laboratory or center-of-mass frame. In particular, the isothermal prescription and its extensions~\cite{Becattini:2021iol,Sheng:2025cjk,Palermo:2025imv} are of considerable phenomenological interest~\cite{Palermo:2024tza,Yi:2024kwu,Arslan:2024dwi,Jiang:2026cnz}, since including shear-induced polarization~\cite{Liu:2021uhn,Fu:2021pok,Becattini:2021suc} can more readily reproduce the observed sign of local polarization when temperature gradients are neglected. The lack of a unified treatment, together with the substantial impact of hypersurface choices on physical predictions, makes the hypersurface dependence of local equilibrium a fundamental and widely discussed conceptual problem in heavy-ion physics.

In this work, we formulate the Zubarev response approach for general hypersurface geometries, treating the nondissipative and so-called ``dissipative'' parts of the density operator on an equal footing. As will be illustrated in detail in this work, treating both parts systematically removes the apparent hypersurface dependence from the local equilibrium definition and provides new insight into the physical role of reference frames.

\sect{The formal aspects of hypersurface covariance}
The Zubarev formalism~\cite{zubarev1974nonequilibrium} provides a convenient and widely used framework~\cite{Liu:2021nyg,Becattini:2021iol,Sheng:2024pbw,Zhang:2024mhs,Yang:2024fkn,Buzzegoli:2025zud,She:2024rnx,Tiwari:2024trl,Dey:2024hhc,Harutyunyan:2021rmb,Harutyunyan:2025fgs,Sheng:2025cjk,Palermo:2025imv,Akkelin:2025lgf,Vitiuk:2024cor,Wang:2025mfz} for identifying and treating the hypersurface dependence of local equilibrium. Its density operator is defined as
\begin{align}
	\label{eq_localfull}
	&\hspace{5em}\rho=e^{-\mathcal{B}}/Z, \quad Z=\text{Tr}[e^{-\mathcal{B}}]\\
	\mathcal{B}&=\int \mathrm{d}\Sigma_{\g}T^{\g\l}\beta_{\l}(x')-\int^{\Sigma}_{\Sigma^{\bm0}}d^4x'\pd_\g\beta_\l(x') T^{\g\l}(x'),\no\\
	\mathcal{B}&=\int \mathrm{d}\Sigma_{\g}T^{\g\l}\beta_{\l}(x')-\int^{\Sigma}d^4x'\pd_\g\beta_\l(x') T^{\g\l}(x')e^{-\e \mathfrak{t}(x')}\no
\end{align}
where $T^{\g\l}$ is the energy-momentum tensor and $\beta_\mu=u_\mu/T$ the inverse-temperature four-vector, with $\beta=1/T$.
Here $\Sigma$ is the hypersurface on which the local-equilibrium part is defined, and $\Sigma^{\bm0}$ is the initial hypersurface. Integration by parts separates the instantaneous local-equilibrium (nondissipative) term on $\Sigma$ from a history integral of $T^{\g\l}(x')\pd_\g\beta_\l(x')$. Since time averaging (coarse-graining) can give rise to effective dissipation, this history term is conventionally called the ``dissipative'' part.

The third line implements Zubarev's central idea: averaging over past initial times to eliminate explicit dependence on a particular choice of initial conditions~\cite{zubarev1974nonequilibrium}, where the factor $e^{-\e\mathfrak{t}(x')}$ is an adiabatic damping regulator that suppresses contributions from the remote past. 
In the covariant formulation used here, this also removes the explicit reference to a particular initial hypersurface $\Sigma^{\bm0}$. However, the nondissipative part of $\mathcal{B}$, routinely included in calculations, still depends explicitly on the choice of hypersurface. For the two hypersurfaces $\Sigma$ and $\tilde{\Sigma}$, we define the difference as
\begin{align}
	&\D\mathcal{B}^{\text{nd}}=\int \mathrm{d}\Sigma_{\mu}T^{\mu\nu}\beta_{\nu}-\int \mathrm{d}\tilde{\Sigma}_{\mu}T^{\mu\nu}\beta_{\nu}.
\end{align}

The dissipative part is often argued to be small or effectively neglected~\cite{Becattini:2021suc,Sheng:2024pbw,Zhang:2024mhs,Sheng:2025cjk,Liu:2021nyg,Yang:2024fkn}, allowing the above hypersurface dependence of the non-dissipative part to enter the final results.
However, this commonly neglected dissipative part also differs between the two hypersurfaces:
\begin{align}
\D\mathcal{B}^{\text{d}}=-\int^{\Sigma}_{\tilde{\Sigma}}d^4x'\pd_\mu\beta_\n T^{\mu\nu}
\end{align}
At the formal level, energy-momentum conservation and Gauss's theorem imply an exact cancellation between these two differences:
\begin{align}
	\label{eq_gausszero}
	\D\mathcal{B}=\D\mathcal{B}^{\text{nd}}+\D\mathcal{B}^{\text{d}}=0\rightarrow\rho(\Sigma)= \rho(\tilde{\Sigma})
\end{align}
This establishes hypersurface covariance for Zubarev's original construction, given in the third line of Eq.~(\ref{eq_localfull}), in the adiabatic limit $\e\to0^+$.  In formulations keeping a specified initial hypersurface $\Sigma^{\bm0}$~\cite{Becattini:2018duy,Becattini:2019dxo,Becattini:2025rqp}, the remaining dependence on $\Sigma^{\bm0}$ reflects the initial-state preparation. In that approach, one might expect that, after the operator is consistently evolved from the initial-state hypersurface with appropriate averaging, the dependence on the initial hypersurface would eventually be removed, following the same logic as Zubarev's original idea.


We next give an explicit realization of this formal cancellation through a polarization calculation using Zubarev's original density operator.

\sect{Zubarev response for general hypersurfaces}
We first generalize our Zubarev response approach, formulated in the medium rest frame in Ref.~\cite{Li:2025pef}, to general hypersurfaces. To treat the nondissipative and dissipative contributions on an equal footing, we express the three-dimensional hypersurface integral as a four-dimensional integral using the standard delta function representation~\cite{Coleman:2011xi}, so that Eq.~(\ref{eq_localfull}) becomes:
\begin{align}
	\label{eq_localK}
	&\mathcal{B}=\int \mathrm{d}^4x'K_{\l\g}(x')T^{\l\g}(x'),\,K_{\l \g}=K^{\text{nd}}_{\l\g}+K^{\text{d}}_{\l\g}\no\\	
	&\hspace{3em}K^{\text{nd}}_{\l\g}(x')=\d(\phi(x'))\pd_\l\phi(x')\beta_\g(x')\no\\
	&\hspace{3em}K^{\text{d}}_{\l\g}(x')=-\theta(-\phi(x'))\pd_\l\beta_\g(x')e^{-\e \mathfrak{t}\cdot(x-x')}
\end{align}
The factor $e^{-\e \mathfrak{t}\cdot(x-x')}$ introduces infinitesimal dissipation, with $\mathfrak{t}$ specifying the time direction of the dissipative process. In the common case~\cite{Kovtun:2012rj}, this direction is aligned with the fluid four-velocity, which defines the medium rest frame. We therefore set $\mathfrak{t}=u$ below and suppress this factor in the notation when no ambiguity arises. Other choices are discussed later.


As a concrete example, we revisit the axial-current calculation that revealed shear-induced polarization~\cite{Liu:2021uhn,Becattini:2021suc} to illustrate the use of Eq.~(\ref{eq_localK}) for general hypersurfaces. Following the conventions of Ref.~\cite{Li:2025pef}, we obtain
\begin{align}
\label{eq_J5_G}
&\hspace{-0.5em}\mathcal{J}_5^\mu(x_{-},x_{+})=\int d^4x'K_{\g\l}(x')\int^{\bar{t}'}_{-\infty}d\bar{t}^{\,\prime\prime} G^{\mu\g\l}\no\\
&\hspace{-0.5em}G^{\mu\g\l}=\frac{-i}{\b}\text{Tr}_{Z_0}\Big\{[J_5^\mu(x_{-},x_{+}),T^{\g\l}(\bar{t}^{\,\prime\prime},\bar{x}^{\prime\m}_\perp)]e^{-\beta u\cdot P}\Big\}.
\end{align}
In this lab-frame expression, $\text{Tr}_{Z_0}$ is the trace normalized by $Z_0=\text{Tr}[e^{-\beta u\cdot P}]$ and $J^\mu_5$ is the point-split axial-current operator. We use $x_\pm=x\pm y/2$ and $k/k'=p\pm q/2$. Since imaginary-time evolution is generated by $u\cdot P$, we decompose $x'$ as $\bar{t}'=u\cdot x'$ and $\bar{x}^{\prime\m}_{\perp}=\bar{\D}^{\m\a}x'_\a$, where $\bar{\D}^{\m\a}=g^{\m\a}-u^\m u^\a$. 
The notation $T^{\g\l}(\bar{t}^{\,\prime\prime},\bar{x}'_\perp)$ denotes the operator evaluated at $u\bar{t}^{\,\prime\prime}+\bar{x}'_\perp$, with $\bar{t}^{\,\prime\prime}$ an auxiliary real-time variable.
Expanding around the equilibrium operator $e^{-\beta u\cdot P}$ introduces dependence on the medium four-velocity $u$, illustrating how physical frame vectors enter the description of a medium. Our goal is not to construct a frame-free theory, but to establish a covariant theory that preserves physical frame dependence while removing spurious frame dependence  from improper approximations.

Equation~(\ref{eq_J5_G}) can be evaluated by explicitly tracking the medium four-velocity~\cite{Weldon:1982aq} and decomposing the propagator momenta into components parallel and orthogonal to $u$, with the relevant $i\e$ prescriptions also expressed in the medium rest frame. The covariance established in Ref.~\cite{Weldon:1982aq} ensures that this treatment is equivalent to performing the calculation in the medium rest frame and then transforming the result to the frame of interest. Using this equivalence, we define the transformation as
\begin{align}
\label{eq_transtou}
G^{\mu\g\l}&=\frac{-i}{\beta}\text{Tr}_{Z_0}\Big\{[U^{-1}J_5^\mu U,U^{-1}T^{\g\l}U]U^{-1}e^{-\beta u\cdot P}U\Big\}\no\\
&=\bar{\Lambda}^\m{}_{\m'}\bar{\Lambda}^{\g}{}_{\g'}\bar{\Lambda}^{\l}{}_{\l'}\bar{G}^{\m'\g'\l'}\no\\
\bar{G}^{\m\g\l}&=\frac{-i}{\beta}\text{Tr}_{Z_0}\Big\{[\bar{J}_{5}^{\m} ,\bar{T}^{\g\l}]e^{-\beta \bar{u}\cdot P}\Big\}
\end{align}
Here $U=U(\bar{\Lambda})$ and $\bar{u}=\bar{\Lambda}^{-1}u=(1,\bm0)$. 
In $\bar{G}$ and below, bars denote quantities in the medium rest frame, with $\bar{x}=\bar{\Lambda}^{-1}x$.

Since the free-field correlator has already been evaluated explicitly in Ref.~\cite{Li:2025pef}, we refer to that work for the calculation and focus on two differences relevant here.
First, the Wigner relative coordinate $y$ can have a nonzero time component in the medium rest frame, so the equal-time treatment must be extended. For the nearly free particle-density contribution considered here, we use the on-shell approximation, which introduces an additional phase $e^{-i\ve_{\bar{p}}\bar{y}^0}$ at leading order in gradients. Its Fourier transform gives the on-shell energy constraint, while the analytic continuation in the frequency conjugate to the center-time difference remains unchanged.
Second, for a general hypersurface, the upper limit $\bar{t}'$ in Eq.~(\ref{eq_J5_G}) need not be earlier than the observation time in the medium rest frame. We therefore require the full commutator rather than only its retarded part. This commutator can be obtained from $G_R-G_A$ through the spectral representation. The auxiliary $i\e$ in the retarded Green function of Ref.~\cite{Li:2025pef} can be taken to zero first, independently of the regulator in the Zubarev history average.
After factoring out the tensor structure, we define the scalar spectral factor for the on-shell density-density contribution as
\begin{align}
\rho_G(\bar{\o},\bar{k},\bar{k}')&=-\delta(\bar{\o}-\bar{u}\cdot \bar{q})(n(\ve_{\bar{k}})-n(\ve_{\bar{k}'}))\;\no\\
&=\delta(\bar{\o}-\bar{u}\cdot \bar{q})\bar{u}\cdot \bar{q}\beta n_0(1-n_0), 
\end{align}
where $n_0=n(u\cdot p)$, with the second line expanded to leading order in $q$.
The Green function can be written as
\begin{align}
&\bar{G}^{\m\g\l}(\bar{t}'',\bar{\vx}',\bar{y})=\int \frac{d^3\bar{\vp}d^3\bar{\vq} d\bar{\o}}{(2\pi)^6\ve_{\bar{k}'}\ve_{\bar{k}}} e^{-i\bar{p}\bar{y}+i\bar{\vq }\cdot(\bar{\vx}-\bar{\vx}')-i\bar{\o} (\bar{t} -\bar{t}'')}\no\\
&\hspace{8em}\times\frac{1}{\b}\rho_G(\bar{\omega},\bar{\vk},\bar{\vk}')\bar{p}_\rho \bar{q}_\s \varepsilon^{\r\s\mu (\g}\bar{p}^{\l)}
\end{align}
After integrating over $\bar{t}''$, we obtain the following expression, denoting the time-integrated kernel by $\bar{G}^{\m\g\l}(\bar{x}',\bar{y})$:
\begin{align}
\label{eq_g_med}
&\bar{G}^{\m\g\l}(\bar{x}',\bar{y})=\int \frac{d^3\bar{\vp}d^3\bar{\vq}d\bar{\o}}{(2\pi)^6\ve_{\bar{\vk}'}\ve_{\bar{\vk}}}e^{-i\bar{p}\bar{y}-i\bar{q}_{\o}\cdot(\bar{x}-\bar{x}')} n_0(1-n_0)\no\\
&\hspace{6em}\times(-i)\delta(\bar{\o}-\bar{u}\cdot \bar{q})\bar{p}_\rho \bar{q}_\s\varepsilon^{\r\s\mu (\g}\bar{p}^{\l)}
\end{align}
where $\bar{q}_{\o}=(\bar{\o},\bar{\vq})$. 
The Lorentz matrices in Eq.~(\ref{eq_transtou}) contracted with $K_{\mu\nu}$ in Eq.~(\ref{eq_localK}) transform it to the medium rest frame. Since $d^4x'$ is Lorentz invariant, we change variables to $\bar{x}'=\bar{\Lambda}^{-1}x'$, giving
%
\begin{align}
	\label{J5_G_rest}
	&\bar{\mathcal{J}}_5^\mu(\bar{y},\bar{x})=\int d^4\bar{x}'\bar{K}_{\g\l}(\bar{x}') \bar{G}^{\mu\g\l}(\bar{x}',\bar{y})
\end{align}
Note that $\mathcal{J}_5=\bar{\Lambda}\bar{\mathcal{J}}_5$.
To obtain the leading-derivative result, we use the local tangent-plane approximation $\phi(x')=-l\cdot(x-x')$, with $\pd_\g\phi(x')=l_\g$, where $l$ is the future-directed unit vector normal to $\Sigma$. In the local hypersurface frame, variables are dressed with tildes and $\tilde{l}=(1,\bm0)$, so $\theta(-\tilde{\phi}(\tilde{x}'))=\theta(\tilde{l}\cdot(\tilde{x}-\tilde{x}'))=\theta(\tilde{t}-\tilde{t}')$.
  

We next transform to the hypersurface frame, with $\bar{\mathcal{J}}=\tilde{\Lambda}\tilde{\mathcal{J}}$. Taking $\mathfrak{t}=u$ as an example, we obtain
\begin{align}
	\label{J5_G_hyper}
	&\tilde{\mathcal{J}}_5^\mu(\tilde{y},\tilde{x})=\int d^4\tilde{x}'\tilde{K}_{\g\l}(\tilde{x}')\tilde{G}^{\mu\g\l}(\tilde{y},\tilde{x}-\tilde{x}')\\
	&\td{K}^{\text{nd}}_{\l\g}=\d(\tilde{t}-\tilde{t}')\td{l}_\l[-(\tilde{x}-\tilde{x}')_\perp]^\xi\td{\pd}_\xi\td{\b}_\g(\td{x})\no\\
	&\td{K}^{\text{d}}_{\l\g}=-\theta(\tilde{t}-\tilde{t}')e^{-\e( \tilde{u}_0(\tilde{t}-\tilde{t}')+\tilde{u}_\perp\cdot (\tilde{x}-\tilde{x}')_\perp)}
	\td{\pd}_\l\td{\b}_\g(\td{x}')
\end{align}
We can express the coordinate factors as momentum derivatives acting on the Fourier amplitudes under the integral. To simplify the notation, we define the scalar kernel $K^{\text{d}}$ by $K^{\text{d}}_{\l\g}=K^{\text{d}}\pd_\l\beta_\g$.
\begin{align}
	\label{eq_Kndd_pdq}
	&\td{K}^{\text{nd}}_{\l\g}=\d(\tilde{t}-\tilde{t}')\td{l}_\l[-(\frac{\pd}{i\pd \td{q}^{\perp}_\xi})]\td{\pd}_\xi\td{\b}_\g,\no\\
	&\td{K}^{\text{d}}=-\theta(\tilde{t}-\tilde{t}')\exp[-\e( \tilde{u}_0(\tilde{t}-\tilde{t}')+\tilde{u}^{\perp}_{\xi} (\frac{\pd}{i\pd \td{q}^{\perp}_\xi}))]
\end{align}

For the nondissipative part, we first perform the frequency integral to obtain the on-shell expression, then express the remaining quantities in the hypersurface frame and evaluate the momentum derivatives. Here $\ve_{\td{p}}=l\cdot p$ and
$\td{q}_0=-\td{p}_\perp\cdot\td{q}_\perp/\ve_{\td{p}}$
at leading order.
Using the Wigner-transform convention of Ref.~\cite{Li:2025pef}, we obtain
\begin{align}
\label{eq_np}
&\tilde{\mathcal{J}}_{\text{nd},5}^\mu=\frac{1}{2}n_0(1-n_0)\varepsilon^{\mu\r\s \g}(\tilde{p}_\rho\tilde{\pd}^{\perp}_\s\td{\b}_\g-2\tilde{p}_\g\tilde{l}_\r\frac{\tilde{p}^{\l}}{\ve_{\tilde{p}}}\tilde{\pd}^{\perp}_{(\s}\td{\b}_{\l)})\no\\
&\equiv	\frac{1}{2}n_0(1-n_0)\varepsilon^{\mu\r\s \g}(\tilde{p}_\rho\tilde{\pd}_\s\td{\b}_\g-2\tilde{p}_\g\tilde{l}_\r\frac{\tilde{p}^{\l}}{\ve_{\tilde{p}}}\tilde{\pd}_{(\s}\td{\b}_{\l)})		
\end{align}
The longitudinal derivatives in the second line cancel exactly between the vorticity and shear terms, leaving explicit $l^\mu$ dependence only in the shear term. 
With $l^\mu$ chosen as the CM time direction~\cite{Becattini:2021iol} or the hypersurface normal $n^\mu$~\cite{Sheng:2025cjk}, Eq.~(\ref{eq_np}) matches the corresponding expressions within our setup, including the isothermal case.



However, the last line of Eq.~(\ref{eq_Kndd_pdq}) introduces additional momentum derivatives outside the medium rest frame. Despite the $\e$ factor, their contributions can survive as $\e\to0^+$ after time integration. Integrating over $\tilde{t}'$ and evaluating the derivatives at $\tilde{q}_\perp=0$, we obtain
\begin{align}
	\label{eq_kernel}
	&\int d\tilde{t}'\td{K}^{\text{d}}e^{-i \tilde{q}_0(\tilde{t}-\tilde{t}')}(-i\tilde{q}_\s)\no\\
	&=
	-\exp[-\e\tilde{u}^\perp_{\xi}\cdot (\frac{\pd}{i\pd \tilde{q}^{\perp}_\xi})]\frac{-i\tilde{q}_\s}{i\tilde{q}_0+\tilde{u}_0\e}\no\\
	&=-[\sum_{n=1}\frac{1}{n!}(\tilde{u}^{\perp}_\xi\frac{\e\,\pd}{-i\pd \tilde{q}^{\perp}_\xi})^n][\sum_{m=0}\frac{-i\tilde{q}_\s}{\e\tilde{u}_0}(\frac{-i\tilde{q}_0}{\e\tilde{u}_0})^m]\no\\
	&=-(\tilde{u}^\perp_\s-\frac{\tilde{p}_\perp\cdot \tilde{u}_\perp }{\ve_{\tilde{p}}}\tilde{l}_\s)[\sum_{m=0}\frac{1}{\tilde{u}_0}(\frac{-\tilde{p}_\perp\cdot \tilde{u}_{\perp}}{\ve_{\tilde{p}}\tilde{u}_0})^m]\no\\
	&=(\frac{\tilde{l}_\s}{\ve_{\tilde{p}}}-\frac{\tilde{u}_\s}{\tilde{u}\cdot\tilde{p}})\ve_{\tilde{p}}
\end{align}
Inserting this kernel into Eq.~(\ref{eq_g_med}), we obtain the contribution from the dissipative part:
\begin{align}
	\label{eq_j5d_final}
\tilde{\mathcal{J}}_{\text{d},5}^\mu=n_0(1-n_0)(\frac{\tilde{l}_\s}{\ve_{\tilde{p}}}-\frac{\tilde{u}_\s}{\tilde{u}\cdot\tilde{p}}) \tilde{p}_\rho\varepsilon^{\mu\s\g\r }\tilde{p}^{\l}\tilde{\pd}_{(\l}\tilde{\beta}_{\g)}
\end{align}
Adding both contributions and transforming back to the lab frame, we recover the hypersurface-covariant result:
\begin{align}
	\label{eq_final}
&\mathcal{J}_{5}^\mu
=\bar{\Lambda}^\m{}_\b\tilde{\Lambda}^\b{}_\a(\tilde{\mathcal{J}}_{\text{nd},5}^\a+\tilde{\mathcal{J}}_{\text{d},5}^\a)\no\\
&=\frac{1}{2}n_0(1-n_0)\varepsilon^{\mu\r\s \g}(p_\rho\pd_\s\beta_\g-2p_\g u_\r\frac{p^{\l}}{u\cdot p}\pd_{(\s}\beta_{\l)})\no\\
&\equiv
\frac{1}{2}n_0(1-n_0)\varepsilon^{\mu\r\s \g}(p_\rho\pd^{\perp}_\s\beta_\g-2p_\g u_\r\frac{p^{\l}}{u\cdot p}\pd^{\perp}_{(\s}\beta_{\l)})
\end{align}
In the last line, $\perp$ denotes projection orthogonal to $u$. The choice $\mathfrak{t}=u$ gives the result above. For a general timelike direction $\mathfrak{t}$, the same calculation yields
{\thickmuskip=1.0mu 
	\medmuskip=1.0mu  
\begin{align}
	\label{eq_final_witht}
\hspace{-0.5em}	\mathcal{J}_{5}^\mu(\mathfrak{t})=	\frac{1}{2}n_0(1-n_0)\varepsilon^{\mu\r\s \g}(p_\rho\pd^{\perp}_\s\beta_\g-2p_\g \mathfrak{t}_\r\frac{p^{\l}}{\mathfrak{t}\cdot p}\pd^{\perp}_{(\s}\beta_{\l)})
\end{align}
}Here $\perp$ denotes projection orthogonal to $\mathfrak{t}$.
The result is independent of the hypersurface but depends on the time direction $\mathfrak{t}$ of dissipative relaxation. For ordinary hydrodynamic relaxation in an isotropic medium with no independent drift direction, $\mathfrak{t}$ coincides with $u$ at leading order~\cite{Kovtun:2012rj}. At the order considered, the medium determines $\mathfrak{t}$, while the hypersurface can be varied without changing the result.
\begin{align}
\mathfrak{t}=u, \quad\text{leading-order hydrodynamics}
\end{align}

One might be concerned that the results above still require specifying $\mathfrak{t}$ with adiabatic damping. 
We will show how to obtain the same result with a more physical damping model, which also allows us to study memory effects. To proceed, we first replace the delta function in Eq.~(\ref{eq_g_med}) with a Lorentzian:
\begin{align}
	\delta(\bar{\o}-\bar{u}\cdot \bar{q})\rightarrow\frac{1}{\pi}\frac{\G}{(\bar{\o}-\bar{u}\cdot \bar{q})^2+\G^2}
\end{align}
This  profile describes exponential damping of the density-density correlation in the medium with rate $\G$, which is twice the quasiparticle damping rate and encodes interaction effects at leading skeleton order~\cite{Li:2025pef}.
Evaluating the frequency integral by residues gives $e^{-\G|u\cdot(x-x')|}$. 
Defining $\d\tilde{x}_u^\perp\equiv\tilde{u}_\perp\cdot(\tilde{x}-\tilde{x}')_{\perp}$, $\theta^\pm\equiv\theta(\pm\d\tilde{x}_u^\perp)$, and $\theta_\tau^\pm\equiv\theta[\pm u\cdot(x-x')]$, the kernel is 
\begin{align}
\td{K}^d&=-[\theta^{+}_\tau(\theta^{+}+\theta^{-})e^{-\G(\tilde{u}\cdot (\tilde{x}-\tilde{x}'))}+\theta^{-}_\tau\theta^{-}e^{\G(\tilde{u}\cdot (\tilde{x}-\tilde{x}'))}].\no
\end{align}
Integrating $\td{K}^{\text{d}}e^{-i\tilde{q}_0(\tilde{t}-\tilde{t}')}(-i\tilde{q}_\s)$ over $\tilde{t}'\leq\tilde{t}$ gives
\begin{align}
	\label{eq_dampmain}
	&-[\theta^{+}\exp[-\G\tilde{u}^\perp_{\xi}\cdot (\frac{\pd}{i\pd \tilde{q}^{\perp}_\xi})]+\theta^{-}e^{i\tilde{q}_0\d \tilde{x}^\perp_u/\tilde{u}_0}]\frac{-i\tilde{q}_\s}{i\tilde{q}_0+\tilde{u}_0\G}\no\\
	&-\theta^{-}[\exp[\G\tilde{u}^\perp_{\xi}\cdot (\frac{\pd}{i\pd \tilde{q}^{\perp}_\xi})]-e^{i\tilde{q}_0\d \tilde{x}^\perp_u/\tilde{u}_0}]\frac{-i\tilde{q}_\s}{i\tilde{q}_0-\tilde{u}_0\G}\no\\
	&=\exp[-\G\tilde{u}^\perp_{\xi}\cdot (\frac{\pd}{i\pd \tilde{q}^{\perp}_\xi})]\frac{(-)(-i\tilde{q}_\s)}{i\tilde{q}_0+\tilde{u}_0\G}|_{q=0}\no\\
	&=(\frac{\tilde{l}_\s}{\ve_{\tilde{p}}}-\frac{\tilde{u}_\s}{\tilde{u}\cdot\tilde{p}})\ve_{\tilde{p}}
\end{align}
Here the momentum derivatives are evaluated before taking $q\to0$ at fixed $\G>0$. Factors of $\G$ from the derivatives are compensated by inverse powers of $\G$ from the time integral, leaving a finite contribution. We thus recover Eq.~(\ref{eq_final}), previously obtained with $\mathfrak{t}=u$, without introducing an auxiliary damping factor.

Indeed, this Lorentzian approximation to the density-density correlator yields an analytic expression for the decay of initial-hypersurface memory~\cite{Becattini:2018duy,Becattini:2019dxo,Becattini:2025rqp}. For $\Sigma^{\bm0}$ defined by $\phi_0=\mf{t}_0-\mf{l}\cdot(x-x')=0$, the additional volume integral in its rest frame gives (see End Matter):
\begin{align}
\hspace{-0.5em}\int d^4 x'\theta(-\phi_{0})\td{K}^di\tilde{q}_\s=-\ve_{\td{p}}(\frac{\td{\mf{l}}_\s}{\ve_{\td{p}}}-\frac{\td{u}_\s}{\td{u}\cdot \td{p}})e^{-\G (\td{u}\cdot\td{p}/\ve_{\td{p}})\mf{t}_0}
\end{align}
where the derivation is similar to that for the $\theta^+$ term in the first line of Eq.~(\ref{eq_dampmain}), with the additional phase factor $e^{-i\td{q}_0\mf{t}_0}$ requiring careful treatment. 
With the notation in Eq.~(\ref{eq_final_witht}), 
 $ \mathcal{J}^\mu_5$ relaxes exponentially from the initial-hypersurface result toward the medium-rest-frame result:
\begin{align}
\mathcal{J}^\mu_5=\mathcal{J}^\mu_5(u)+(\mathcal{J}^\mu_5(\mf{l})-\mathcal{J}^\mu_5(u))e^{-( u\cdot p/\mf{l}\cdot p)\G \mf{t}_0}
\end{align}
This formula resolves the disagreement over hypersurface dependence by describing both initial-hypersurface memory~\cite{Becattini:2018duy,Becattini:2019dxo,Becattini:2025rqp} at short times and the medium-rest-frame  result~\cite{Liu:2021uhn,Yi:2021ryh} at long times in a single relaxation process.




\sect{Toward hypersurface covariance at higher orders}
Gauss's theorem implies the formal cancellation in Eq.~(\ref{eq_gausszero}), demonstrated explicitly above at leading gradient order. Here we outline a practical route to realizing it order by order for general hypersurface geometries, starting from a general operator $\mathcal{B}$:
\begin{align}
	&\mathcal{B}=\int \mathrm{d}^4x'\big[\d(\phi(x'))\pd_\l\phi(x') X_\g(x')\\
	&\qquad-\theta(-\phi(x'))e^{-\e \mathfrak{t}\cdot(x-x') }\pd_\l X_\g(x')\big] Y^{\l\g}(x')\no
\end{align}
where $X_\g$ is a source field and $Y^{\l\g}$ is a conserved current satisfying $\pd_\l Y^{\l\g}=0$. For a general operator $O^b(a,x)$, expanding the density operator expresses its expectation value in terms of correlation functions with insertions of $Y^{\l\g}$. Here $b$ denotes Lorentz indices and $a$ collects other arguments. The linear-response contribution is
\begin{align}
	&\<O^b(a,x)\>_h=\int\mathrm{d}^4x'\big[\d(\phi_h(x'))\pd_\l\phi_h(x')X_\g(x')\no\\
	 &-\theta(-\phi_h(x'))e^{-\e \mathfrak{t}(x') }\pd_\l X_\g(x')\big]G^{b\l\g}(a,x,x')
\end{align}
Here $h\in[0,1]$ parametrizes a smooth deformation between the two hypersurfaces.
\begin{align}
	\label{eq_D_all}
	&\<O^b(a,x)\>_{h=1}-\<O^b(a,x)\>_{h=0}\\
	&=\int d^4x'G^{b\l\g}(a,x,x')[\D K^{nd}_{\l\g}(x')+\D K^{d}_{\l\g}(x')]\no
\end{align}	
where the nondissipative contribution is
\begin{align}
	\label{eq_Dknd}
	\D K^{nd}_{\l\g}=&\big(\d(\phi_{1}(x'))\pd_\l\phi_1(x')\no\\
	&-\d(\phi_{0}(x'))\pd_\l\phi_0(x')\big)X_\g(x')
\end{align}
and the dissipative contribution is
\begin{align}
	\label{eq_Dkd}
		&\D K^{d}_{\l\g}=\int^{1}_0 dh\,\delta(\phi_h(x'))\frac{d\phi_h(x')}{dh}e^{-\e \mathfrak{t}(x') }\pd_\l X_\g(x')\\
		&=\int^{1}_0 dh [d(\frac{d\theta(-\phi_h(x'))}{d\phi}\pd_\l \phi_h )/dh] X_\g(x')\no\\
		&=-\int^{1}_0 dh \frac{d[\d(\phi_h(x'))\pd_\l \phi_h] }{dh} X_\g(x')=-\D K^{nd}_{\l\g}\no
\end{align}
The integration-by-parts relations above are understood after contraction with $G^{b\l\g}(a,x,x')$ and integration over $x'$, with vanishing boundary terms. They require $\pd'_\l G^{b\l\g}(a,x,x')=0$, so current conservation must be preserved by the approximation used to evaluate the correlator. The last two lines display the formal cancellation with the adiabatic regulator temporarily set to unity.

The first line can be used directly in explicit calculations. For two planar hypersurfaces with normals $l^{\bm0}$ and $l^{\bm1}$, we define $\dot{l}\equiv dl/dh$ and obtain the following difference in the dissipative response at leading order in gradients:
\begin{align}
&\D\mathcal{J}^\mu_{\text{d},5}=n_0(1-n_0)p_\r \D l_\s\ve^{\r\s\m(\g}p^{\l)}\pd_\g \b_\l\\
&\D l_\s= \int^1_0dh\,\Big(\frac{\dot{l}_{\s}}{l\cdot p}-
\frac{\dot{l}\cdot p}{(l\cdot p)^2}
l_\s\Big)=\frac{l^{\bm1}_\s}{l^{\bm1}\cdot p}-\frac{l^{\bm0}_\s}{l^{\bm0}\cdot p}
\end{align}
Choosing $l^{\bm0}=u$ and $l^{\bm1}=l$ reproduces Eq.~(\ref{eq_j5d_final}).

\sect{Summary}
In this work, we establish a Zubarev response theory for general hypersurface geometries, treating the nondissipative and so-called ``dissipative'' parts of the density operator on an equal footing. We develop a resummation technique that reveals a finite nondissipative contribution from the ``dissipative part" when the hypersurface normal differs from the direction of dissipative evolution. At leading order in gradients, this contribution exactly cancels the hypersurface dependence of the local-equilibrium polarization, restoring hypersurface covariance.


With adiabatic damping, the remaining frame dependence has a physical origin: it enters through the medium four-velocity $u$ and the time direction $\mathfrak{t}$ of the dissipative process. For ordinary hydrodynamic relaxation in an isotropic medium, $\mathfrak{t}$ and $u$ coincide at leading order, and different hypersurface choices recover the medium-rest-frame result.  

For finite damping, introducing a Lorentzian spectral width into the density-density correlator reproduces the medium-rest-frame result in the long-time limit, without an auxiliary damping factor or an independently specified $\mathfrak{t}$. At finite times, it yields an explicit relaxation formula describing the exponential decay of initial-hypersurface memory, which resolves the disagreement over hypersurface dependence by connecting the short-time memory of the initial hypersurface with the long-time medium-rest-frame result in a single relaxation process.

Hypersurface prescriptions substantially affect polarization predictions~\cite{Fu:2021pok,Becattini:2021iol,Yi:2024kwu}. Resolving this ambiguity clarifies the physical interpretation of these predictions and provides a consistent basis for applying the Zubarev approach to a broad range of observables in heavy-ion collisions. In the future, we will investigate memory effects in microscopic interacting models beyond leading skeleton order, extend the response calculation to higher orders in gradients, and explore the phenomenological implications of memory effects and higher-order corrections.

\acknowledgments
\textbf{Acknowledgments}
We are grateful to Shan-Ping Wu, Francesco Becattini, Matteo Buzzegoli, Andrea Palermo, and Xin-Li Sheng for helpful discussions. SL is supported by NSFC Grant No. 12205090 and the Fundamental Research Funds for the Central Universities. 

\bibliography{ref}
\section{End matter}
In this section, we present a more detailed derivation of the initial-hypersurface memory effect. The initial hypersurface is defined by $\phi_0=\mf{t}_0-\mf{l}\cdot(x-x')=0$. The volume integral can then be written as
\begin{align}
	&-\int d^4 x'\theta(-\phi_{0})\td{K}^d(-i\tilde{q}_\s)e^{-i\td{q}\cdot(\td{x}-\td{x}')}\no\\
	&=\int d\td{t}'  d^3 \td{x}'_{\perp}\theta(-\phi_{0})e^{-\Gamma |\td{u}_0(\td{t}-\td{t}')+\td{u}_\perp\cdot(\td{x}-\td{x}')_{\perp}|}(-i\td{q}_\s)\no\\
	 &\quad\times\exp[{i\td{q}_\perp\cdot[-(\td{x}-\td{x}')_{\perp}+\frac{\td{p}_\perp}{\ve_{\td{p}}}(\td{t}-\td{t}')]}]
\end{align}
Here $\perp$ denotes projection orthogonal to $l$. The initial-hypersurface condition $\phi_0=0$ reads $\td{\mf{l}}_0(\td{t}-\td{t}')+\td{\mf{l}}_\perp\cdot(\td{x}-\td{x}')_\perp=\mf{t}_0$. Defining $\td{\mf{f}}=\td{u}-(\td{u}_0/\td{\mf{l}}_0)\td{\mf{l}}$ and $\theta^\pm=\theta(\pm[\td{\mf{f}}_\perp\cdot(\td{x}-\td{x}')_\perp+(\td{u}_0/\td{\mf{l}}_0)\mf{t}_0])$, we first perform the time integral and obtain the following expression, with the remaining spatial integral and the common Fourier factor $e^{-i\td{q}_\perp\cdot(\td{x}-\td{x}')_\perp}$ left implicit:
\begin{align}
&\big[\theta^{+}e^{i\td{q}_0(\td{\mf{l}}^{\perp}\cdot(\td{x}-\tilde{x}')_{\perp}- \mf{t}_0)/\td{\mf{l}}_0}e^{-\G \td{u}_0\mf{t}_0/\td{\mf{l}}_0}e^{-\G \td{\mf{f}}^\perp\cdot(\td{x}-\tilde{x}')_{\perp}}\no\\
&\quad+\theta^{-}e^{i\td{q}_0(\td{u}^{\perp}/\td{u}_0)\cdot(\td{x}-\tilde{x}')_{\perp}}\big]\frac{-i\td{q}_\s}{i\td{q}_0+\td{u}_0\G}\no\\
&+\theta^{-}\big[e^{i\td{q}_0(\td{\mf{l}}^{\perp}\cdot(\td{x}-\tilde{x}')_{\perp}- \mf{t}_0)/\td{\mf{l}}_0}e^{\G \td{u}_0\mf{t}_0/\td{\mf{l}}_0}e^{\G \td{\mf{f}}^\perp\cdot(\td{x}-\tilde{x}')_{\perp}}\no\\
&\quad-e^{i\td{q}_0(\td{u}^{\perp}/\td{u}_0)\cdot(\td{x}-\tilde{x}')_{\perp}}\big]\frac{-i\td{q}_\s}{i\td{q}_0-\td{u}_0\G}
\end{align}


Due to the covariance, we can simplify the expansion in the initial-hypersurface rest frame, where $\td{\mf{l}}=(1,\bm0)$ and the
$\td{\mf{l}}_\perp$-dependent spatial phase equals unity. Below, tildes and $\perp$ refer to this frame, with $\ve_{\td{p}}=\mf{l}\cdot p$. For sufficiently large $\mf{t}_0$, one may assume that the physical integration region of $x'$ lies entirely in the $\theta^+$ branch. Under this assumption, local expansion and resummation yield:
\begin{align}
	\label{eq_dampf1}
	&[\exp[-\G(\tilde{u}^\perp_{\xi})\cdot (\frac{\pd}{i\pd \tilde{q}^{\perp}_\xi})]]\frac{-i\tilde{q}_\s}{i\tilde{q}_0+\tilde{u}_0\G}e^{-i\td{q}_0 \mf{t}_0}e^{-\G (\td{u}\cdot\td{\mf{l}})\mf{t}_0} \;\big|_{\td{q}_\perp=0}\no\\
	&=\sum_{n=0}^{\infty}\sum_{m=0}^{n}(\tilde{u}^\perp_\s-\frac{\tilde{p}_\perp\cdot \tilde{u}_\perp }{\ve_{\tilde{p}}}\tilde{\mf{l}}_\s)\frac{1}{\tilde{u}_0}(\frac{-\tilde{p}_\perp\cdot \tilde{u}_{\perp}}{\ve_{\tilde{p}}\tilde{u}_0})^m\no\\ &\quad\times\frac{(-\G\mf{t}_0\td{p}_{\perp}\cdot\td{u}/\ve_{\td{p}})^{n-m}}{(n-m)!}e^{-\G (\td{u}\cdot\td{\mf{l}})\mf{t}_0}\no\\	
	&=-\ve_{\td{p}}(\frac{\td{\mf{l}}_\s}{\ve_{\td{p}}}-\frac{\td{u}_\s}{\td{u}\cdot \td{p}})e^{-\G (\td{u}\cdot\td{p}/\ve_{\td{p}})\mf{t}_0}
\end{align}
Reordering the double sum and summing over $n-m$ first gives $e^{-\G(\td{p}_\perp\cdot\td{u}_\perp/\ve_{\td{p}})\mf{t}_0}$. Combining it with the original exponential yields the damping factor in the last line.

To avoid the assumption that restricts the spatial integration to the \(\theta^+\) branch, we present an alternative derivation. 
Still in the initial-hypersurface frame, we first integrate over $\td{q}_\perp$ in $-(2\pi)^{-3}\int d^3\td{q}_\perp\int d^4\td{x}'e^{-i\td{q}\cdot(\td{x}-\td{x}')}\theta(-\phi_0)\td{K}^d(-i\tilde{q}_\s)$ to obtain a delta function in coordinate space:
\begin{align}
	&-\int_{-\infty}^{\td{t}-\mf{t}_0}d\td{t}'d^3\td{x}'_{\perp} \d((\td{x}-\td{x}')_{\perp}-\frac{\td{p}_\perp}{\ve_{\td{p}}}(\td{t}-\td{t}'))\no\\ &\quad\times(\td{\mf{l}}_{\s}\frac{\td{p}_{\perp}}{\ve_{\td{p}}}\cdot \pd_{\perp} -\pd_{\s}^{\perp})e^{-\Gamma (\td{u}_0(\td{t}-\td{t}')+\td{u}_\perp\cdot(\td{x}-\td{x}')_{\perp})}
\end{align}
Here we remove $|\cdots|$, since the delta function and the integration limits ensure that its argument is positive. Taking the derivatives gives:
\begin{align}
	&-
	\int_{-\infty}^{\td{t}-\mf{t}_0}d\td{t}' d^3\td{x}'_{\perp}\d((\td{x}-\td{x}')_{\perp}-\frac{\td{p}_\perp}{\ve_{\td{p}}}(\td{t}-\td{t}'))\no\\ &\quad\times \G(\td{\mf{l}}_{\s}\frac{\td{p}_{\perp}\cdot \td{u}_\perp}{\ve_{\td{p}}} -\td{u}_{\s}^{\perp}) e^{-\Gamma (\td{u}_0(\td{t}-\td{t}')+\td{u}_\perp\cdot(\td{x}-\td{x}')_{\perp})} 
\end{align}
Integrating over $\td{x}'_\perp$ gives:
\begin{align}
	&-\G(\td{\mf{l}}_{\s}\frac{\td{p}_{\perp}\cdot \td{u}_\perp}{\ve_{\td{p}}} -\td{u}_{\s}^{\perp})
	\int_{-\infty}^{\td{t}-\mf{t}_0}d\td{t}' e^{-\Gamma (\frac{\td{u}\cdot \td{p}}{\ve_{\td{p}}}(\td{t}-\td{t}'))}
\end{align}
Performing the remaining time integral gives:
\begin{align}
	\label{eq_dampf2}
	&-\G(\td{\mf{l}}_{\s}\frac{\td{p}_{\perp}\cdot \td{u}_\perp}{\ve_{\td{p}}} -\td{u}_{\s}^{\perp})\frac{\ve_{\td{p}}}{\td{u}\cdot \td{p}\G}e^{- \frac{\td{u}\cdot \td{p}}{\ve_{\td{p}}}\Gamma \mf{t}_0}\no\\
	&=-\ve_{\td{p}}(\frac{\td{\mf{l}}_\s}{\ve_{\td{p}}}-\frac{\td{u}_\s}{\td{u}\cdot \td{p}})e^{-\G (\td{u}\cdot\td{p}/\ve_{\td{p}})\mf{t}_0}
\end{align}
Eqs.~(\ref{eq_dampf1}) and~(\ref{eq_dampf2}) yield the same kernel and hence the same full axial current, which can be expressed in the lab frame as
\begin{align}
	\mathcal{J}^\mu_5=\mathcal{J}^\mu_5(u)+(\mathcal{J}^\mu_5(\mf{l})-\mathcal{J}^\mu_5(u))e^{-( u\cdot p/\mf{l}\cdot p)\G \mf{t}_0}
\end{align}
where the notation follows the convention in Eq.~(\ref{eq_final_witht}).
\vspace{0mm}


%
%
%
%

\end{document}